# Low lying magnetic states of the mixed valence cobalt ludwigite

M. Matos[a], J. Terra[b], and D. E. Ellis[b,c]

[a]*Departamento de Física, PUC-Rio, Rua Marques de São Vicente, 225, Gávea, Rio de Janeiro, RJ, 22453-970, Brazil.*
*Corresponding author, maria.matos@fis.puc-rio.br (55) 21 3527 1263*
[b]*Centro Brasileiro de Pesquisas Físicas, Rio de Janeiro, RJ, Brazil*
[c]*Northwestern University, Evanston, IL, USA*

**Abstract**

There are two interpretations offered for the different structural and magnetic properties of the mixed valence homo-metallic ludwigites, $Co_3O_2BO_3$ and $Fe_3O_2BO_3$. One of them associates the physical behavior to charge ordering processes among the cations, as is well known in simpler oxides. The other attributes the effects to local pairwise magnetic interactions. Recently first principles calculations in the iron ludwigite have shown that the structural cation dimerization is due to the formation of strong magnetic *dyads* supporting the second model. Here we confirm the dominance of magnetic interactions to explain the absence of dimerization in the cobalt compound. Density functional non-collinear spin calculations are carried out on $Co_3O_2BO_3$ to determine its low temperature magnetic order. Low spin is found on tri-valent cobalt sites, thus preventing the formation of the ferromagnetic *dyad*, the mechanism which favors dimerization in $Fe_3O_2BO_3$. We conclude that the difference between high spin $Fe^{3+}$ and low spin $Co^{3+}$ pairwise interactions is responsible for the observed differences between the two compounds. The pairwise magnetic interactions also explain the difference between the existence of low temperature bulk AF state in the Fe ludwigite and its absence in the Co material.

# 1. Introduction

Oxo-borates of the ludwigite structure have received attention in the last decades because the only two homo-metallic mixed valence compounds of this family, $Co_3O_2BO_3$ and $Fe_3O_2BO_3$, have very distinct physical properties. In the iron ludwigite, a room temperature structural rearrangement at 283K is followed by three magnetic regions appearing at 110K, 70K and 40K[**1-4**]. The atomic rearrangement, mainly a dimerization in $Fe^{2+}$-$Fe^{3+}$ pairs, is associated with the appearance of intermediate valences of Fe. Ludwigite has a flat orthorhombic unit cell containing four octahedral metal sites distributed over two atom *triads*, 424 and 313 (**FIG.1**). The 424 *triad* contains the shortest metal-metal bonds of the structure; in the iron compound, magnetic interaction within 424 are responsible for dimerization. The cobalt ludwigite, by contrast, is structurally stable and has only one magnetic phase which sets in at ~ 40K [**5-8**].

In the Fe ludwigite varied metal charge rearrangements occur over a large temperature range, encompassing as well the magnetic transitions [**4**, **8,9**]. In another iron oxo-borate, the warwickite $Fe_2OBO_3$, an orthorhombic to monoclinic transition occurs at 317K, well above the magnetic ordering temperature 155K[**10**]. The structural transition was well understood as a charge ordering Verwey-like transition. For the Fe ludwigite it was suggested that a charge ordering Peierls-like transition could cause dimerization by opening a gap at the Fermi level[**11**]. The model should in principle predict dimerization in the cobalt ludwigite as well, since both compounds have the same crystalline structure. Tight binding calculations [**12**] showed that the band gap is due to iron-oxygen rather than Fe-Fe interactions, and appears both above and below the transition. So far, direct influence of electronic effects such as charge ordering on the $Fe_3O_2BO_3$ structural transition has not been clearly established.

As an alternative explanation it was suggested that magnetic properties could be involved in the structural transition of the iron ludwigite [**4,13,14**]. The main question to be resolved was the big difference between structural and magnetic transition temperatures. More recently[**15**] a close connection was found between dimerization and magnetism in $Fe_3O_2BO_3$, independently of the transition temperatures being so far apart. First-principles non-collinear spin DFT calculations on the low temperature dimerized phase showed that the 424 *triad* consists in fact of two independent magnetic sub-units, a Fe2-Fe4 *dyad*, and a Fe4 cation. The *dyad* behaves as a robust ferromagnetic dimer, with spin flip energy of ~800meV. The Fe4 cation in the opposite extreme of the *triad* has a magnetic moment with great flexibility to rotate and forms the sub-set of canted spins in this material. The large *dyad* magnetic energy gives support to the idea that local ferromagnetic Fe2-Fe4 *dyads* could exist at higher temperatures, influencing the structural transition at 283K and establishing a direct connection between magnetism and structural instabilities. In addition, canting of Fe4 spins explained the anti-ferromagnetic state observed at low temperatures. A Mössbauer spectroscopy study shows pronounced splitting in isomer shifts indicating differences between the two Fe4 sites at lower temperatures **[9]**, consistent with the DFT result.

The understanding of distinct structural behavior of iron and cobalt ludwigites has motivated much experimental research focusing on the magnetic properties of cobalt in ludwigite. Some important experimental data on the magnetic state of $Co_3O_2BO_3$ are now well established. Its magnetic structure has been mainly characterized as weak ferromagnetic or ferrimagnetic [**6,8,16**], with magnetization ~ 40 emu/g, in contrast with the antiferromagnetic state of $Fe_3O_2BO_3$. Magnetization was found to be highly anisotropic and perpendicular to the short crystal axis $c$ of the orthorhombic unit cell [**8,16**]. Freitas et al. **[17]** performed neutron scattering experiment in $Co_3O_2BO_3$ and the magnetic state was determined, showing the presence of low spin in Co4 and ferromagnetic alignment along $c$ in the 424 triads.

Inverse susceptibility versus temperature measurements led to an imprecise definition of the sign of the Curie-Weiss temperature [**6, 18**] which was found to be one order of magnitude smaller than in Fe ludwigite [**9,18**]. This made it difficult to predict the type of spin alignment in the cobalt system. Calculations based on a model super-exchange spin Hamiltonian [**19**] could not describe the complex magnetic state of the system. Vanishing of magnetization under zero field cooling has been observed [**6, 16**] probably due to the presence of magnetic domains in $Co_3O_2BO_3$. Disappearance of hysteresis curves above ~5K [**8**] suggests that interactions between domains are small.

Mixed cobalt ludwigites were also investigated from the point of view of magnetism [**7, 8,16, 18**-**26**]. In the iron/cobalt ludwigites $Co_2FeO_2BO_3$ [**7,16**] and $Co_{2.25}Fe_{0.75}O_2BO_3$ [**8,19**], the characteristic magnetic transition of $Co_3O_2BO_3$ disappears, giving way to magnetic orderings around 110K and 70K, which are characteristic of the parent iron ludwigite. The transition at 110K was associated to magnetic ordering of $Fe^{3+}$ spins in site 4. In $Co_2FeO_2BO_3$ [**7**] specific heat measurements showed no feature at 70K, indicating an absence of long range magnetic ordering. A simple freezing of magnetic moments of Co was thus assumed at this temperature [**7, 19**]. In manganese substituted $Co_{1.7}Mn_{1.3}O_2BO_3$[**20**], saturation occurs at 41K, roughly the same magnetic transition temperature as the parent $Co_3O_2BO_3$. Mn is distributed over all crystalline sites with Co preference for site 4. Based on differences of hysteresis curves of the mixed and pure compounds and a possible influence of disorder, the magnetic order was interpreted as a spin-glass freezing. Mixing with non-magnetic Ti in $Co_5Ti(O_2BO_3)_2$ [21] leads to magnetization saturation at temperature 19K. Since, as with the Fe mixed compound, no feature was found in specific heat measurements, the magnetic state below 19K was characterized as a spin glass freezing. In $CoMgGaO_2BO_3$ [22], $Co_{2.4}Ga_{0.6}O_2BO_3$ [23] and $Co_{2.88}Cu_{0.12}O_2BO_3$ [24], whose magnetic temperatures were found to be 25K, 37K and 43K, respectively, spin glass freezing ($CoMgGaO_2BO_3$) or ferrimagnetism ($Co_{2.4}Ga_{0.6}O_2BO_3$ and $Co_{2.88}Cu_{0.12}O_2BO_3$) was suggested. An interesting case is the substitution with tin, $Co_5Sn(O_2BO_3)_2$ [26]. So far, it is the only hetero-metallic ludwigite to show a sharp peak in the specific heat versus temperature curve, thus indicating long range magnetic ordering. Note that $Ni_2FeO_2BO_3$ [7] also showed no long range order. In the Sn substituted compound the magnetic transition occurs at 82K, twice as big as that of the pure cobalt ludwigite (~ 41K). The magnetic state was associated to ferrimagnetic ordering. It should be noted that no dimerization transition was found in any of these compounds.

Although these works have considerably improved understanding of the physics of the ludwigites, their magnetic state, in particular that of the pure cobalt ludwigite, remain unknown.

In this paper we investigate the magnetic order of $Co_3O_2BO_3$, by using the first principles non-collinear spin DFT methodology. Our aim is to bring new light on the interplay between magnetism and dimerization in the homo-metallic ludwigite structure and improve understanding on the magnetic order of this compound. Since there are different valences of cobalt in the system, possible different spin states are considered in this investigation, leading to a quantitative description of low-lying magnetic states.

**2. Crystal structure**

$Co_3O_2BO_3$[27] is synthesized in the ludwigite structure which is formed by metal containing edge-sharing oxygen octahedra displayed in corrugated planes, held together by strongly bonded $BO_3$ units. Co has thus octahedral coordination with oxygen. The material has a flat orthorhombic unit cell of *Pbam* symmetry with lattice constants *a*=9.275Å, *b*=12.146Å and *c* = 3.027Å. It is a mixed-valence compound with four formula units per unit cell that can be written as $[(Co^{2+}_2Co^{3+})(O_2BO_3)^{-7}]_4$. The structure is shown in **FIG.1** projected in the *ab* plane. The 4 non-equivalent Co sites form two *triads* 424 ($Co^{3+}4$-$Co^{2+}2$-$Co^{3+}4$) and 313 ($Co^{2+}3$-$Co^{2+}1$-$Co^{2+}3$). The 424 *triad* in the ludwigites is responsible for important physical properties since it has the shortest metal-metal bond, whose value is $d_{Co2\text{-}Co4}$= 2.789Å, in the cobalt compound. In the secondary 313 *triad*, Co-Co bonding is weaker, with $d_{Co1\text{-}Co3}$= 3.384 Å.

The present calculations use the experimental structure of $Co_3O_2BO_3$[27] to define the atomic positions of the unit cell. In order to investigate magnetic structures with periodicity (*a*, *b*, *2c*) **[15]**, the unit cell is duplicated along *c* and has the 72-atom composition $[(Co^{2+}_2Co^{3+})(O_2BO_3)^{-7}]_8$.

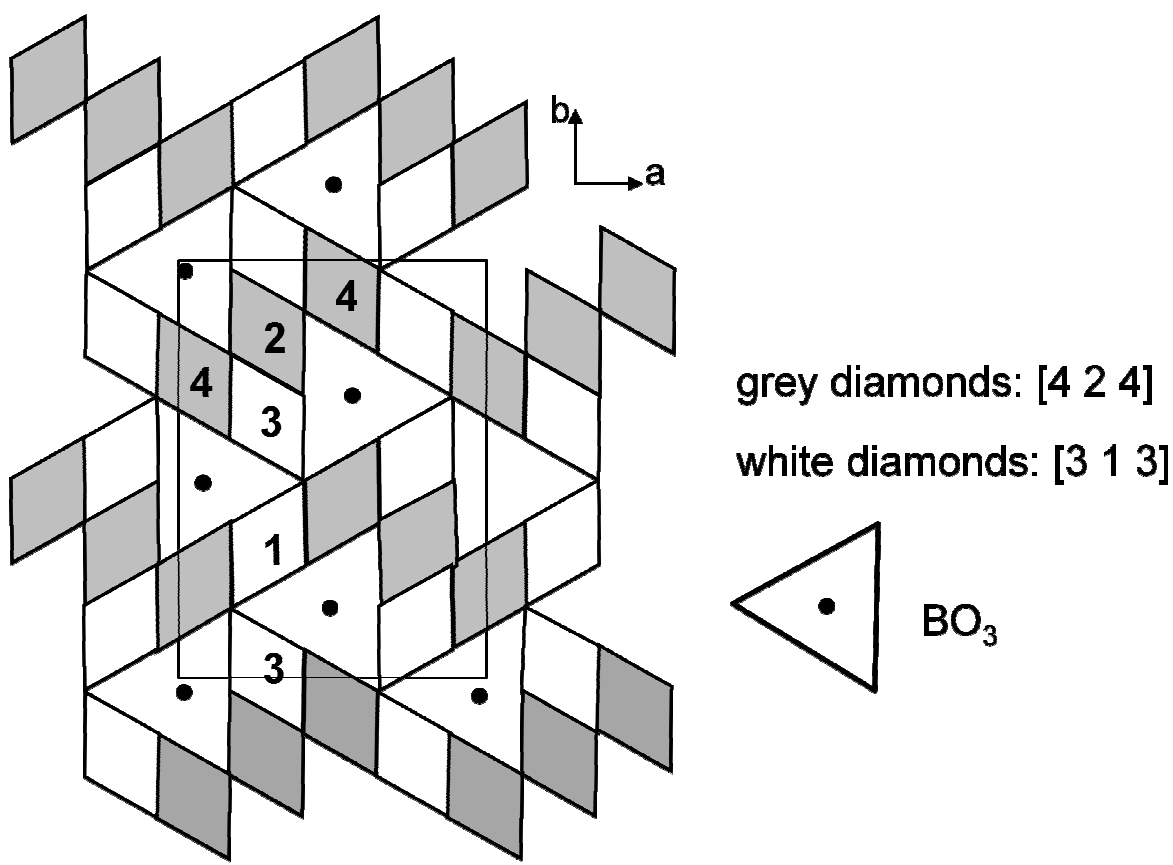

**FIG.1** – The octahedral representation of the ludwigite structure seen along the short orthorhombic axis, *c*. Metal sites define the *triads* 424 and 313. Small black circles: B.

## 3. Details of calculation

The quantum mechanical electronic structure of $Co_3O_2BO_3$ is obtained by using the first principles density functional theory (DFT) in the non-collinear spin-polarized approach of the VASP code [28]. The projector-augmented wave method and a plane-wave basis set were employed using the so-called generalized gradient PAW-GGA PW91[29, 30] approximation to describe the exchange and correlation scheme, with an energy cutoff of 450 eV. The electronic basis set is composed of [Ar] for Co, leaving $3d^8 4s^1$ valence electrons and [He] for both B and O leaving $2s^2 2p^1$ and $2s^2 2p^4$ valence electrons for B and O, respectively. It has been suggested that spin-orbit coupling could be relevant in the description of the electronic structure of the cobalt ludwigite [6], thus we examine these effects by using the spin-orbit methodology implemented in VASP.

For the *initial* spin configuration required in the self-consistent calculation, two kinds of input models (**AF/b** and **F/b**) are analyzed. They are schematically shown in **Table 1**. Both models assume spin orientation in the 424 *triad* to be parallel to the lattice axis *b* and orthogonal to that of *triad* 313; this arrangement was determined from neutron diffraction data [14] and further confirmed by first principles calculations [15] in the low temperature magnetic state of iron ludwigite. The difference between the two models consists of spin ordering along *c* in the 424 *triads*, which is anti-ferromagnetic in spin model **AF/b** and ferromagnetic in spin model **F/b**. The symbol **/b** means that the 424 spins are parallel to the *b* axis. The **AF** *c*-alignment in the 424 triad was shown to provide an excellent description of magnetic properties of $Fe_3O_2BO_3$[15]. The ferromagnetic **F/b** model takes into account the cobalt spin ordering obtained by Freitas et al. [17]. **AF/b** and **F/b** have respectively initial magnetizations of $(M_x, M_y, M_z) = (-12.0, 0.0, 0.0)$ $\mu_B$ and $(-12.0, 44.0, 0.0)$ $\mu_B$ per unit cell (*a*, *b*, *2c*). Initial configurations in which spins of 424 and 313 *triads* point in the same direction were found have unit cell energies too large as compared to arrangements in which 424 and 313 spins are perpendicular, in accordance with neutron scattering results [14]. Formal values of magnetic moments of $Co^{2+}$ (S=3/2) and $Co^{3+}$(S= 2), were initially assumed. They were fully allowed to vary through the spin self-consistent calculation to determine the final spin state of each cation.

To investigate magnetization anisotropy and determine the preferred direction of this vector with respect to the crystalline axis, two additional input models, **AF/a** and **F/a**, were considered. They consist of a 90º rotation of the whole spin systems of models **AF/b** and **F/b**, preserving their relative individual orientations (see **Table 1**). The new input magnetizations are, accordingly, (0.0,-12.0, 0.0)$\mu_B$ (for **AF/a**) and (44.0, -12.0, 0.0)$\mu_B$ (for **F/a**).

We will refer to calculations which consider the non-collinear spin polarized calculations as SPIN and those which include the spin-orbit coupling as LS. Good convergence of the total energy is obtained with the 5x5x6 Monkhorst-Pack k-space grid within ~1 meV as compared with the 4x4x5 grid.

**Table 1** – Idealized initial spin configurations, **AF/b**, **F/b**, **AF/a** and **F/a** with z-components equal zero. The duplication of *triads* in each entry indicates spin arrangement along the *c* axis. **AF** and **F** mean magnetic ordering of 424 spins along *c*; alignment in the 313 triad is always ferromagnetic. /*a* and /*b* mean spin orientation of triads 424 relative to lattice axes *a* and *b*. Non collinear arrangement of each model is evident in the different orientations of 424 and 313 spins.

| model | Initial spin configuration of $Co_3O_2BO_3$ | | |
|---|---|---|---|
| | double cell along *c* | Configuration in the ab plane | |
| | | 424 | 313 |
| **AF/b** | (0,0,1) <br> (0,0,0) | ↑ ↑ ↑  *b*↑ <br> ↓ ↓ ↓ | → ← →  *a*→ <br> → ← → |
| **F/b** | (0,0,1) <br> (0,0,0) | ↑ ↑ ↑  *b*↑ <br> ↑ ↑ ↑ | → ← →  *a*→ <br> → ← → |
| **AF/a** | (0,0,1) <br> (0,0,0) | → → →  *a*→ <br> ← ← ← | ↓ ↑ ↓  *b*↑ <br> ↓ ↑ ↓ |
| **F/a** | (0,0,1) <br> (0,0,0) | → → →  *a*→ <br> → → → | ↓ ↑ ↓  *b*↑ <br> ↓ ↑ ↓ |

## 4. Results

In this section we present results obtained for the calculated electronic and magnetic structure of $Co_3O_2BO_3$. **Table 2** shows the system's total energy and magnetization per unit cell, obtained with the several initial spin configurations described in **Table 1**. For each model, calculations were done using both the SPIN and LS (Spin Orbit Coupling) methodologies.

**Table 2** – Total energy per unit cell E relative to the ground state $E_0$ = -509.292 eV (see text); components ($M_x$, $M_y$, $M_z$) and absolute value (|M|) of the magnetization per unit cell ($a$, $b$, $2c$). Energy in meV, magnetic quantities in $\mu_B$. For comparison, at 40K thermal energy is ~ 3 meV.

| **Non-collinear spin polarized (SPIN)** | | |
|---|---|---|
| | **AF/b** | **F/b** |
| $E-E_0$ (meV) | 352 | 465 |
| $M_{x=a}$ | -10.983 | -12.268 |
| $M_{y=b}$ | -0.170 | 8.713 |
| $M_{z=c}$ | -0.003 | -0.0040 |
| \|M\| | 10.984 | 15.047 |
| | **AF/a** | **F/a** |
| $E-E_0$ (meV) | 347 | 461 |
| $M_{x=a}$ | 0.224 | 9.139 |
| $M_{y=b}$ | -10.501 | -13.131 |
| $M_{z=c}$ | 0.001 | 0.006 |
| \|M\| | 10.503 | 15.998 |
| **Non-collinear with spin orbit (LS)** | | |
| | **AF/b** (first excited state) | **F/b** |
| $E-E_0$ (meV) | 6 | 116 |
| $M_{x=a}$ | -9.802 | -12.696 |
| $M_{y=b}$ | -0.005 | 10.828 |
| $M_{z=c}$ | -0.002 | -0.010 |
| \|M\| | 9.802 | 16.686 |
| | **AF/a** (ground state) | **F/a** |
| $E-E_0$ (meV) | **0** | 104 |
| $M_{x=a}$ | **-0.001** | 9.760 |
| $M_{y=b}$ | **-9.901** | -11.301 |
| $M_{z=c}$ | **0.008** | 0.004 |
| \|M\| | **9.901** | 14.932 |

It is found that the best approximation for the ground state of $Co_3O_2BO_3$ is obtained by using the **AF/a** input model and spin orbit coupling (LS), with the lowest energy given by $E_0$=-509.292 eV. As defined above, in the input model **AF/a** the 424 *triad* spins are parallel to the *a*-axis and align anti-ferromagnetically along *c*. The 313 cobalt spins are parallel to the *b*-axis. For this state, the magnetic moment directions of cobalt in each site are rotated by 90° relative to those of $Fe_3O_2BO_3$ **[15]**.

Naturally a great number of alternative spin states could be considered. For example, modifying the **AF/a** input to make spin *c*-ordering of the 313 *triad* anti-ferromagnetic we obtained $E-E_0$ = 772 meV higher than that found for the ground state, far above kT at the magnetic transition temperature 40K (~3 meV), not shown in **Table 2**.

The histogram of **FIG.2** shows the energy difference $\Delta E = E-E_0$ disposed in decreasing order. It is readily seen that spin-orbit coupling is the most relevant physical feature in determining the ground state and separates the calculations in two distinct groups, SPIN and LS. By taking the average of calculated energies within each group one obtains <E(SPIN)> - <E(LS)>= 346 meV, well above kT. Within each group (SPIN or LS) the **AF** spin alignment of the 424 *triad* remains a preferred arrangement when compared to the ferromagnetic (**F**) one. Their separation, <$E_F$>- <$E_{AF}$> = 113 meV/107 meV , respectively for SPIN/LS cases, is about 35 kT. The average is calculated between **/b** and **/a** spin orientations.

Dependence of the system´s energy E on the orientation of the spin axis relative to the lattice axis is small if compared with the effect produced by spin orbit coupling and by the magnetic ordering of the 424 spins. This can be seen in **FIG.2** by comparing the histogram bars of models **/a** and **/b**. Taking the LS group of calculations, for each kind of 424 spin ordering (**F/** and **AF/**), the LS-energy differences between the two spin to lattice orientations are $\Delta E_F$ =E(**F/b**)- (**F/a**)= 12meV and $\Delta E_{AF}$ = E(**AF/b**)-E(**AF/a**) = E(**AF/b**)-$E_0$= 6 meV. For T = 40K, the ratio of thermal statistical probability between the ground state (**AF/a**) and the next excited state (**AF/b**) is given by exp(- $\Delta E_{AF}$ /kT) = exp (-6/3) ~ 0.14**.** This small ratio suggests that different magnetization directions could be associated to the presence of magnetic domains of $Co_3O_2BO_3$**.** The association of magnetic domains with magnetic anisotropy in ludwigites is corroborated by experiment. In $Co_3O_2BO_3$, which exhibits magnetic anisotropy, magnetization was shown to vanish under zero field cooling [6,8,16], indicating the presence of magnetic domains [6]. In the mixed ludwigite $Co_{2.4}Ga_{0.6}O_2BO_3$ [23], magnetization was found to be finite under zero field cooling. Consistently, no preferred magnetic direction was found in the *ab* plane. As expected, non-collinear calculations without spin-orbit coupling are found to give negligible energy differences under rotation of the spin axis. Absolute value of magnetization varies 3% with orientation as seen in **Table 2**, giving an indication of the level of precision of the calculation.

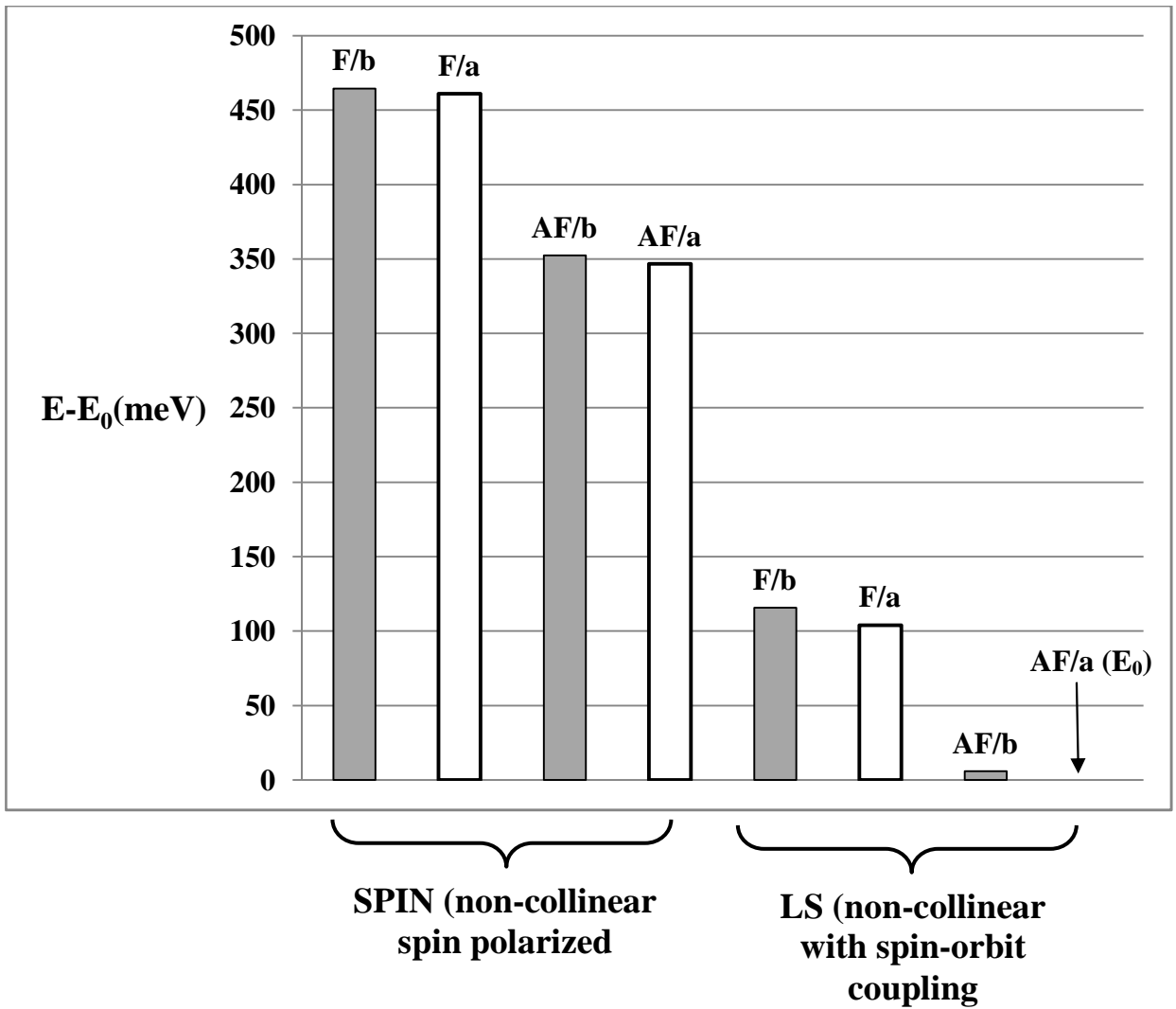

**FIG.2** – Histogram of total energy per unit cell of $Co_3O_2BO_3$ with respect to the ground state **AF/a (LS)**, in decreasing order. Grey(white) bars: 424 spins parallel to *b(a)*. Energies in meV.

By comparing SPIN and LS results in **Table 2**, it is noted that |M| depends little on spin orbit coupling. In fact, by defining, for each kind of ordering (**AF/** or **F/**), the relative difference $\delta_{LS-SPIN} =2[<|M|(LS)>-<|M|(SPIN)>]/[<|M|(LS)> + <|M|(SPIN)]>$, where $<|M|>$ is the average value obtained with the two spin to lattice directions, **/a** and **/b**, we obtain $\delta_{LS-SPIN}$ = -0.083 and -0.039 for **AF** and **F** alignments respectively. These quantities are nearly 1/1000 of the respective absolute magnetization values, which, as shown, vary from 9.7 $\mu_B$ to 16.5 $\mu_B$. This constitutes evidence of negligible orbital contribution to magnetization in cobalt ludwigite. Our results show similarity with $Co_3O_4$ [32] where an orbital contribution to the magnetic moment of $Co^{2+}$ as small as ~ 1/10 of the spin only value was found through neutron diffraction.

Our calculations confirm the low spin state of Co4, found by Freitas et al. [17]. Contrary to $Fe_3O_2BO_3$, in which the magnetic moment of Fe4 plays an important role, Co4 is in a low-spin state. The calculated magnetic moment $\mu$(Co4) ~ 0.2 $\mu_B$ is comparable to the induced magnetic moments obtained for the non-magnetic ions, boron and oxygen, and is one order of magnitude smaller than in the other cobalt sites, Co1, Co2 and Co3, of ~2.3 $\mu_B$ so that one could consider Co4 to be non-magnetic.

This is the most important difference observed so far between the two homo-metallic ludwigites and gives strong support to the role of magnetism in the structural properties of the two compounds. In $Fe_3O_2BO_3$[15], the ferromagnetic *dyad* found in the dimerized structure has a spin flip energy large enough to be sustained in higher temperatures, so that local Fe2-Fe4 ferromagnetic dimers could exist, influencing dimerization. In $Co_3O_2BO_3$ this mechanism is hindered due to demagnetization of Co4, thus preventing dimerization.

Atomic magnetic moments obtained for the ground state are schematically represented in **FIG.3**. For each metal site, the atomic magnetic moment varies little, ~ 0.01 $\mu_B$, among the calculations of **Table 2**. The calculated absolute value of magnetization |M| for the ground state (**AF/a, LS**) of 9.943$\mu_B$/unit cell $\cong$ 25.9 emu/g is in close agreement with the experimental value of ~ 35 emu/g [6]. It is essentially parallel to the *b*-axis, in accordance with experiment [**8,16**]. As the 424 *triad* gives zero spin contribution to magnetization its value is due to the 313 *triads*, in spite of the anti-ferromagnetic order between Co1 and Co3 spins. Bulk magnetization is a result of spin imbalance: there are twice as much Co3 as Co1 (see **FIG. 3**). In the Fe ludwigite, the anti-ferromagnetic low temperature state (M ~ 0) is achieved through canting of one of the Fe4 spins, which cancels out the uncompensated magnetization of *triads* 313[15]. In the cobalt ludwigite no canting is found, due to demagnetized Co4.

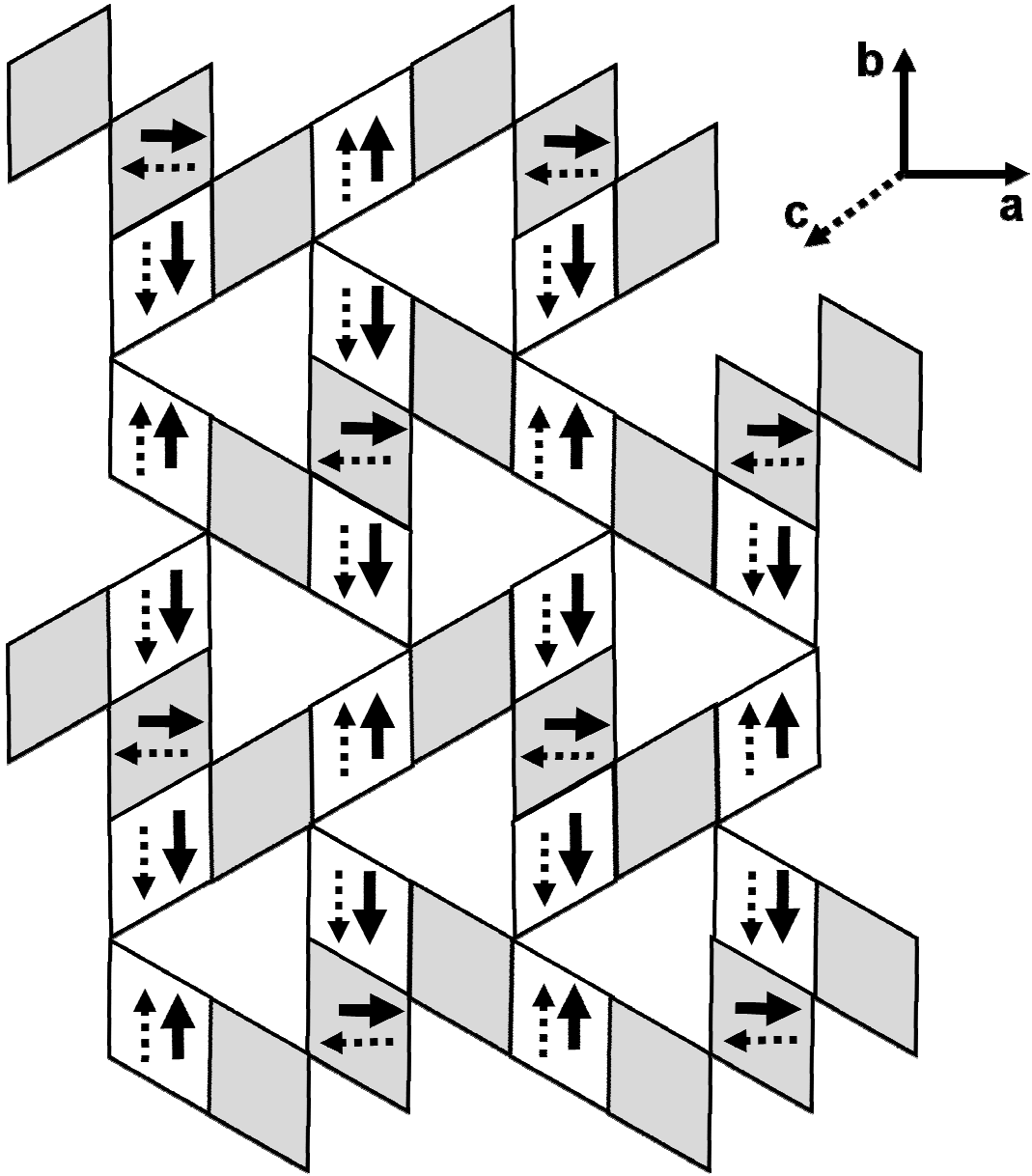

**FIG.3** – The cobalt magnetic moments distribution for the ground state of $Co_3O_2BO_3$ in the double unit cell (*a,b,2c*). Demagnetization of Co4 is represented by absent arrows in site 4. For all sites, $\mu_z < 0.1$ $\mu_B$. Grey diamonds : 424; white diamonds: 313. Dashed arrows: atom in upper cell; solid arrows: atoms in lower cell.

Demagnetization of Co cations has been found in another cobalt oxide, $Co_3O_4$, both through neutron scattering [32] and DFT calculations[34]; it is associated to the $Co^{3+}$ octahedral sites of this normal spinel. From empirical estimates based on bond valence sums in the cobalt ludwigite [6] one gets for the oxidation number of Co4 the value 2.725 while Co1, Co2 and Co3 have oxidation numbers 1.913, 2.056 and 1.977, respectively. This indicates Co4 as a nearly tri-valent cation, while other cobalt sites could be interpreted as di-valent. From this point of view, demagnetization of $Co4^{3+}$ has a close analogy to magnetism in $Co_3O_4$. The structural complexity of the ludwigite, however, leads to net bulk magnetization which is not observed in the oxide.

Using the Pbam primitive unit cell with 4 formula units in a magnetic symmetry analysis, Freitas et al. [17] necessarily obtained ferromagnetic alignment of $Co^{2+}$ in site 2. We were then motivated to perform calculations taking for initial input the spin configuration obtained by these authors. We obtain a total energy of 106 meV (~30 kT for T=40K) above the ground state, comparable with our result for the low lying excited state given by the **F/a** model (see **Table 2**), which considers a ferromagnetic alignment of Co2. In an early study of $Fe_3O_2BO_3$ **[34]**, neutron scattering data were indexed in a magnetic cell doubled along *c* and anti-ferromagnetic ordering was found in the 424 *triad*. We thus decided to perform a magnetic symmetry analysis in $Co_3O_2BO_3$, by means of the technique described by Bertaut **[35]**, using the BASIREPS software **[37]**. For the Pbnm space group with the unit cell doubled along *c*, with 8 formula units, compatible with the magnetic cell duplication, there exist eight real irreducible representations for the little group Gk associated with k=0, Γ1- Γ8. The magnetic arrangement obtained in the present study of the 313 triad belongs to the Γ7 representation whereas that of triad 424 belongs to Γ4. This confirms the anti-ferromagnetic alignment of Co2. The use of *c*-doubled cell in the magnetic analysis thus provides more freedom to adjust the spin configuration of $Co_3O_2BO_3$. Magnetization of the our obtained ground state (**AF/a** of Table 2) is 26 emu/g. Our result for the Freitas et al. input model is 59 emu/g. These values are to be compared with the experimental finding of ~ 40 emu/g **[6]**. Possibly the ferromagnetic alignment proposed by Freitas et al. **[17]** corresponds to a meta stable state of the cobalt ludwigite.

Regarding the absence of dimerization transition in other mixed metal ludwigites, we point out that non-magnetic ions substitutions in the 424 *triad* prevent the formation of the ferromagnetic *dyad*, the physical mechanism of dimerization. It is the case of Ti, and Sn mixed cobalt ludwigites, where substituted ions are located in site 4 [21,26]. In the Ga/Co compound $Co_{2.4}Ga_{0.6}O_2BO_3$ [23] , non-magnetic Ga occupies sites 2 and 4 so a magnetic 2-4 *dyad* cannot be formed. Although in $CoMgGaO_2BO_3$ [22], Mg and Ga sites were not determined, it can be expected that Ga and Mg show preference for sites 4 and 2, respectively, usually associated with tri- and di-valent cations, and this prevents the formation of 2-4 *dyads* as well. Cu occupation site was not determined in $Co_{2.88}Cu_{0.12}O_2BO_3$, nor was its valence, but the small content of copper ions and its low magnetic moment would make it unlikely to find strong magnetic dimers in this compound.

In the mixed Co/Fe[7,8], Ni/Fe[7] and Co/Mn[20] ludwigites, there are 2-4 *hetero-metallic* pairs, Co2-Fe4, Ni2-Fe4 and Co2-Mn4, so that ferromagnetic *dyads* could in principle exist. However, to obtain dimerization electron hopping must be significant; this effect was observed in the pure iron ludwigite, where Mössbauer spectroscopy revealed the presence of intermediate valence in the Fe4-Fe2-Fe4 *triad* [4,9]. However, in the iron mixed compounds the amount of Fe2-Fe4 pairs is small, since site 2 is mainly filled with cobalt. Indeed, attempts to synthesize $Co_{2-x}Fe_xO_2BO_3$ with x>1 did not succeed[8]. It is not clear whether a significant amount of hopping exists in *hetero-metallic* pairs.

**5. Discussion and Conclusion**

In summary, we have found that the differences between the magnetic structures of $Co_3O_2BO_3$ and $Fe_3O_2BO_3$[15] are related to the extreme sites of the 424 metal *triad*. Demagnetization of Co4 is the key feature, since it blocks both the formation of a ferromagnetic Co2-Co4 *dyad* and spin canting in the other Co4. The robust *dyad* found in $Fe_3O_2BO_3$ influences dimerization at higher temperatures by favoring electron hopping in Fe2-Fe4 pairs. And spin canting of the other Fe4, leads to negligible low temperature magnetization in $Fe_3O_2BO_3$. None of these effects are found in $Co_3O_2BO_3$.

We found spin orbit coupling to be irrelevant to determine the magnitude of magnetization, indicating negligible contribution of orbital magnetic moments. However, LS coupling was found to be important in determining low-lying energy states. Calculations showed an important difference between this compound and the other homo-metallic ludwigite, $Fe_3O_2BO_3$, related to the trivalent crystalline site 4, at the edges of the short bonded 424 *triad*. We found that $Co^{3+}$ has negligible magnetic moment, which prevents the formation of a cobalt *dyad*, a strongly interacting ferromagnetic dimer which in iron ludwigite influences the structural transition. Predicted demagnetization of $Co^{3+}$, blocking the structural transition in $Co_3O_2BO_3$, gives theoretical support to the connection between magnetism and structural transition in the homo-metallic ludwigites. As a further consequence we found that spin canting of site 4 in $Fe_3O_2BO_3$, responsible for its low temperature anti-ferromagnetic state **[15]**, is absent in $Co_3O_2BO_3$ due to the small value of its site 4 moment. We found that the 424 *triad* does not contribute to bulk magnetization in $Co_3O_2BO_3$ due to the antiferromagnetic arrangement along *c*; magnetization of this material is controlled by the 313 *triad*.

More research is necessary to determine the magnetic structure and electron-hopping properties of the hetero-metallic ludwigites. This would provide a more complete understanding of the connection between magnetism and structural stability in the ludwigite material in general.